\documentclass{aa}

\usepackage{appendix}
\usepackage[english]{babel}
\usepackage[babel=true]{csquotes}

\usepackage{amsmath}
\usepackage{geometry} 
\usepackage{graphicx} 
\usepackage{fancyhdr}
\usepackage[amssymb]{SIunits}
\usepackage{amsfonts}
\usepackage[version=3]{mhchem}
\usepackage{sectsty}
\usepackage{numprint}
\usepackage{wasysym}
\usepackage{multirow}
\usepackage{fancyhdr}
\usepackage{placeins}
\usepackage{pdflscape}
\usepackage{rotating}
\usepackage{natbib}
\bibpunct{(}{)}{;}{a}{}{,} 

\makeatother

\begin{document}

\title{The different origins of high- and low-ionization broad emission lines revealed by gravitational microlensing in the Einstein cross\thanks{Based on observations made with the ESO-VLT, Paranal, Chile; Proposals 076.B-0197 and 076.B-0607 (PI: Courbin).}}
\titlerunning{Microlensing effect on high- and low-ionization lines in Q2237+0305}

\author{ L.~Braibant\inst{1}, D.~Hutsem\'ekers\inst{1}, D. Sluse\inst{1}, T. Anguita\inst{2}$^,$\inst{3}}
\authorrunning{Braibant et al.}

\institute{
\inst{1} Institut d’Astrophysique, Universit\'{e} de Li\`{e}ge, All\'{e}e du 6 Août 17, B5c, 4000 Li\`{e}ge, Belgium \\
\inst{2} Departamento de Ciencias Fisicas, Universidad Andres Bello, Av. Republica 252, Santiago, Chile \\
\inst{3} Millennium Institute of Astrophysics, Chile
}

\abstract{
We investigate the kinematics and ionization structure of the broad emission line region of the gravitationally lensed quasar QSO2237+0305 (the Einstein cross) using differential microlensing in the high- and low-ionization broad emission lines. We combine visible and near-infrared spectra of the four images of the lensed quasar and detect a large-amplitude microlensing effect distorting the high-ionization CIV and low-ionization H$\alpha$ line profiles in image A. While microlensing only magnifies the red wing of the Balmer line, it symmetrically magnifies the wings of the CIV emission line. Given that the same microlensing pattern magnifies both the high- and low-ionization broad emission line regions, these dissimilar distortions of the line profiles suggest that the high- and low-ionization regions are governed by different kinematics. Since this quasar is likely viewed at intermediate inclination, we argue that the differential magnification of the blue and red wings of H$\alpha$ favors a flattened, virialized, low-ionization region whereas the symmetric microlensing effect measured in CIV can be reproduced by an emission line formed in a polar wind, without the need of fine-tuned caustic configurations.
}

\maketitle

\section{Introduction}

\begin{figure}[t]
\centering
\includegraphics[width=0.5\textwidth]{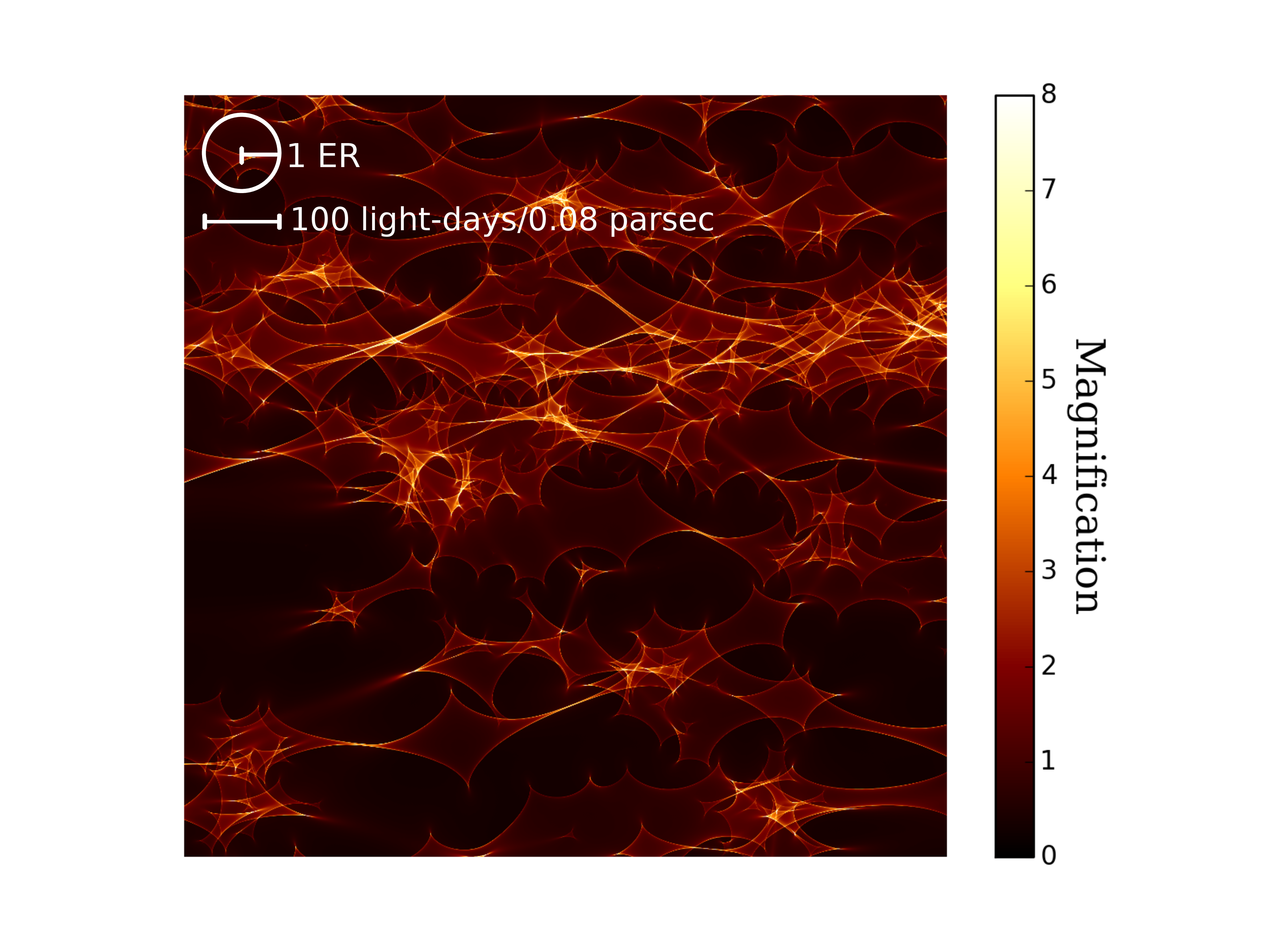}
\caption{20$\times$20 Einstein radii (ER) microlensing magnification map computed for QSO2237+0305A with a convergence factor $\kappa = 0.394$, a shear $\gamma = 0.395$, and a null fraction of continuously distributed matter, using the \texttt{microlens} code \citep{1999Wambsganss}. Caustics are narrow high-magnification features that delineate amplification and de-amplification regions. A circle with a radius of one ER is plotted in the upper left corner.}
\label{fig:mumap}
\end{figure}

Gravitational microlensing is a powerful tool to study the innermost regions of quasars. The dense field of stars in the lensing galaxy produces a microlensing magnification pattern in the source plane, which is constituted of narrow highly-magnifying caustics separated by large weakly magnifying or de-magnifying regions (Fig.~\ref{fig:mumap}). Significant microlensing magnification occurs on the scale of the Einstein radius of the microlenses, corresponding to a few dozen light days for a typical gravitational lens. The caustics can thus resolve the accretion disk and/or sample the broad emission line region (BLR) of the quasar. Due to the cosmological distances of quasars, the only observable effect is the amplification of the continuum and/or the part of the line profile originating from the magnified region of the BLR (\citealp{1986Kayser}, \citealp{2010SchmidtWambsganss}).

Measuring the magnification caused by microlensing in the different spectral components of the quasar spectrum hence provides crucial constraints on the (relative) extensions and positions of their emission regions. In particular, the BLR stratification and ionization structure can be retrieved by comparing the microlensing magnification measured in emission lines with different ionization degrees. Moreover, the magnification of only a part of the BLR results in line profile distortions, for instance displacement of the line centroid. The deformations of the line profile depend on the caustic pattern, on the one hand, and on the BLR geometry and kinematics, on the other hand (e.g., \citealp{2012Sluse}). Information about the BLR spatial and velocity structure can then be inferred from the measurement of the microlensing signal through the line profile.


The gravitational lens system QSO2237+0305 \citep{1985Huchra} has been known for years to be a priviledged laboratory for microlensing studies. The very short time delays, of the order of one day (e.g., \citealp{2006Vakulik}), indeed ease the interpretation of the spectral differences observed between the quasar lensed images in terms of microlensing-induced differences rather than intrinsic variations. This cosmic mirage, also named \enquote{Einstein cross} because of its shape, consists of a $z_s=1.695$ quasar gravitationally lensed into four images separated by about $1.5''$ and arranged in a crosslike pattern around a bright $z_l=0.0394$ barred, Sab lensing galaxy \citep{1988Yee}. 
This system is known to show microlensing-induced temporal and spectral variations (\citealp{1989Irwin}, \citealp{1996Lewis} and references therein). In addition, \citet{2011Sluse} and \citet{2011ODowd} have detected differential microlensing through the velocity structure of the high-ionization carbon line. This lensed quasar therefore appears particularly well suited for investigating the structure of the BLR using gravitational microlensing.


In this work, we analyze near-infrared and visible spectra of the four lensed quasar images, obtained in October 2005 (Sect.~\ref{sec:data}). These combined spectra cover a broad wavelength range that includes the \ce{CIV} $\lambda 1549$, \ce{CIII]} $\lambda 1909$, \ce{MgII} $\lambda 2798$, and \ce{H \alpha} $\lambda 6565$ lines. We present a detailed study of the microlensing effect in image A, i.e. the image that shows the largest differential magnification through the CIV and H$\alpha$ line profiles at that epoch (Sect.~\ref{sec:measure_micro}). We aim to disentangle the part of the quasar spectrum which is microlensed and to build a full picture of how microlensing is distorting the quasar spectrum in the rest-frame UV~(Sect.~\ref{sec:MmD_CIV}) and optical (Sect.~\ref{sec:MmD_Ha}). The constraints on the BLR spatial, velocity and ionization structure set by the microlensing effect detected in image A are discussed in Sect.~\ref{sec:disc-BLRconstrain} and compared to models described in the literature (Sect.~\ref{sec:disc-BLRmodel}). Comparison of the microlensing signal in the high- and low-ionization lines is discussed in Sect.~\ref{sec:blr-ionization-structure}. Conclusions and perspectives are presented in Sect.~\ref{sec:conclusions}.

\section{Data collection and reduction}
\label{sec:data}

\subsection{Near-infrared spectra}

From the ESO archive, we retrieved near-infrared spectra of the quadruply-imaged QSO2237+0305 that was obtained in October 2005 with the integral field spectrograph SINFONI mounted on the Yepun telescope (UT4) of the Very Large Telescope (VLT). Observations were performed using the $3'' \times 3''$ field-of-view (FOV) with $0.1''$ spatial resolution and the H-band grism, whose spectral coverage goes from $1.45$ to $\unit{1.85}{\micro\meter}$ and thus includes the $\ce{H \alpha}$ broad emission line. The H-band grism provides a spectral resolving power around $3\,000$, which corresponds to a spectral resolution of about $\unit{100}{\kilo\meter\per\second}$. The target covers a large part of SINFONI FOV so that 600-second exposures of the target and infrared sky were acquired by nodding the telescope. Table~\ref{tab:sinfo_obs_conditions} (in the appendix) lists the observing conditions.

The ESORex SINFONI pipeline (version 2.5.2) is used to perform flat-fielding, distortion correction, wavelength calibration and sky subtraction, and build 3D-cubes made of monochromatic images of the FOV. Cosmic rays are removed from each monochromatic frame with the \texttt{la{\_}cosmic} procedure \citep{2001VanDokkum}. Non-uniform illumination of the SINFONI FOV is empirically corrected using 3D-cubes of near-infrared sky emission. 3D-cubes of infrared sky emission are built separately for that purpose. All available sky observations are stacked over wavelengths to obtain an image of the FOV illumination. Illumination is found to significantly drop in slitlets \#8-10 but appears constant within a given slitlet. A mean illumination is therefore computed for each slitlet. These mean slitlet illuminations are then normalized by their median and used to rectify the illumination variations over the different slitlets composing the SINFONI FOV.

As in \citet{2014Braibant}, the spectra of the four quasar images are extracted from each 3D-cube by fitting a simplified model of the lensed system to each monochromatic FOV with a modified MPFIT package \citep{2009Markwardt}. The quasar images are modeled with identical 2D Moffat point spread functions (PSFs) whose relative positions are fixed by \citet{2006Kochanek} astrometry, while a PSF-convolved de Vaucouleurs profile is used for the bulge of the lensing galaxy \citep{1988Yee}\footnote{As validation, we used the de Vaucouleurs' model described in \citet{2008Anguita} to perform the spectral extraction and obtained identical results.}. We neglect the disk component of the spiral galaxy because it is expected to be about ten thousand times fainter than the galactic bulge and its brightness is expected to vary by less than 20\% over the SINFONI small FOV. Our model includes a constant background, which is supposed to account for both the residual sky and the diffuse galactic disk emission.

To avoid being trapped in a local minimum, we carefully estimate initial conditions of the model components. First, an image of the H$\alpha$ emission of the gravitationally lensed system, from which the lensing galaxy is essentially absent, is obtained after subtraction of a linear continuum component estimated in two continuum intervals aside of the emission line. This high signal-to-noise picture of the quasar images is used to assess the PSF shape and the system position on the IFU. Second, the spectrum of each quasar image is estimated from the 3D-cube as the spectrum corresponding to the spatial pixel that contains the PSF peak and used to remove the quasar emission from the 3D-cube. The spectrum of the lensing galaxy is then assessed from that residual 3D-cube, which contains the deblended galactic emission. We finally fit the PSFs, de Vaucouleurs' galactic bulge and constant background altogether. The adjusted model is used to iteratively improve the spectral extraction.

Each extracted spectrum is corrected for atmospheric extinction and instrumental response by dividing it by a normalized, blackbody-corrected, telluric standard star spectrum, acquired close in time and at comparable airmass. The H-band magnitudes of the telluric standard stars are listed in the 2MASS Point Source Catalog so that the flux calibration can be performed. Since we are only interested in the spectral distortions caused by gravitational microlensing, we corrected for the dust extinction in the lensing galaxy following \citet{2008Eigenbroda}: while images A and B were found to be free of significant reddening, images C and D were corrected using a Cardelli law with $(A_V,R_V)=(0.2\, \text{mag},3.1)$. 

The error on the spectral flux densities is estimated using the ratios between multiple spectra of the same quasar image obtained on October 29 and 30 2005, the two consecutive observing nights during which most observations were taken and that benefit from good seeing and weather conditions. In each wavelength bin, the relative error on the monochromatic flux density is taken as the standard deviation of the ratios computed between all possible pairs of spectra of a given quasar image:
\begin{center}
$\sigma (F_{X}^{\lambda}) / F_{X}^{\lambda} \simeq \text{stdev} ( F_{X,\text{epoch }i}^{\lambda} / F_{X, \text{epoch }j}^{\lambda} \hspace{1mm} , \hspace{1mm} \forall i\ne j ) / \sqrt{2}$
\end{center}
with $X=A,B,C,D$.

To increase the signal-to-noise, we computed mean near-infrared spectra of the quasar images (top right panel of Fig.~\ref{fig:show_spectra}). We discarded the observations made during the cloudy night of October 26 2005, as well as those acquired with seeing conditions worse than $0.6''$ (see Table~\ref{tab:sinfo_obs_conditions} in the appendix). The spectra of quasar images whose peak is located on slitlets \#8 to 10 were also excluded. We checked that the spectra selected to compute the mean spectrum of each image were consistent with each other.

\subsection{Visible spectra}

Visible spectra of QSO2237+0305 were acquired with the FOcal Reducer and low dispersion Spectrograph (FORS1) mounted on Unit Telescope \# 2 of the VLT under program ID 076.B-0197 (PI: Courbin). A detailed description of the observation strategy and data reduction can be found in \citet{2008Eigenbroda}. For our analysis, we solely consider the spectra acquired on October 11 2005.\footnote{Visible spectra were also acquired on October 1st and October 21st 2005, simultaneously with the SINFONI observations, but they are found to deviate from the OGLE lightcurve (\citealp{2008Eigenbroda}, \citealp{2011Sluse}) and are thus discarded.} These spectra are illustrated in the top left panel of Fig.~\ref{fig:show_spectra}. The FORS1 spectra cover the $[\unit{3800}{\angstrom},\unit{8500}{\angstrom}]$ wavelength range, which contains the \ce{CIV}, \ce{CIII]} and \ce{MgII} emission lines. 


\section{Microlensing in QSO2237+0305 lensed images}
\label{sec:measure_micro}

\begin{figure*}
\begin{center}
\includegraphics[width=0.95\textwidth]{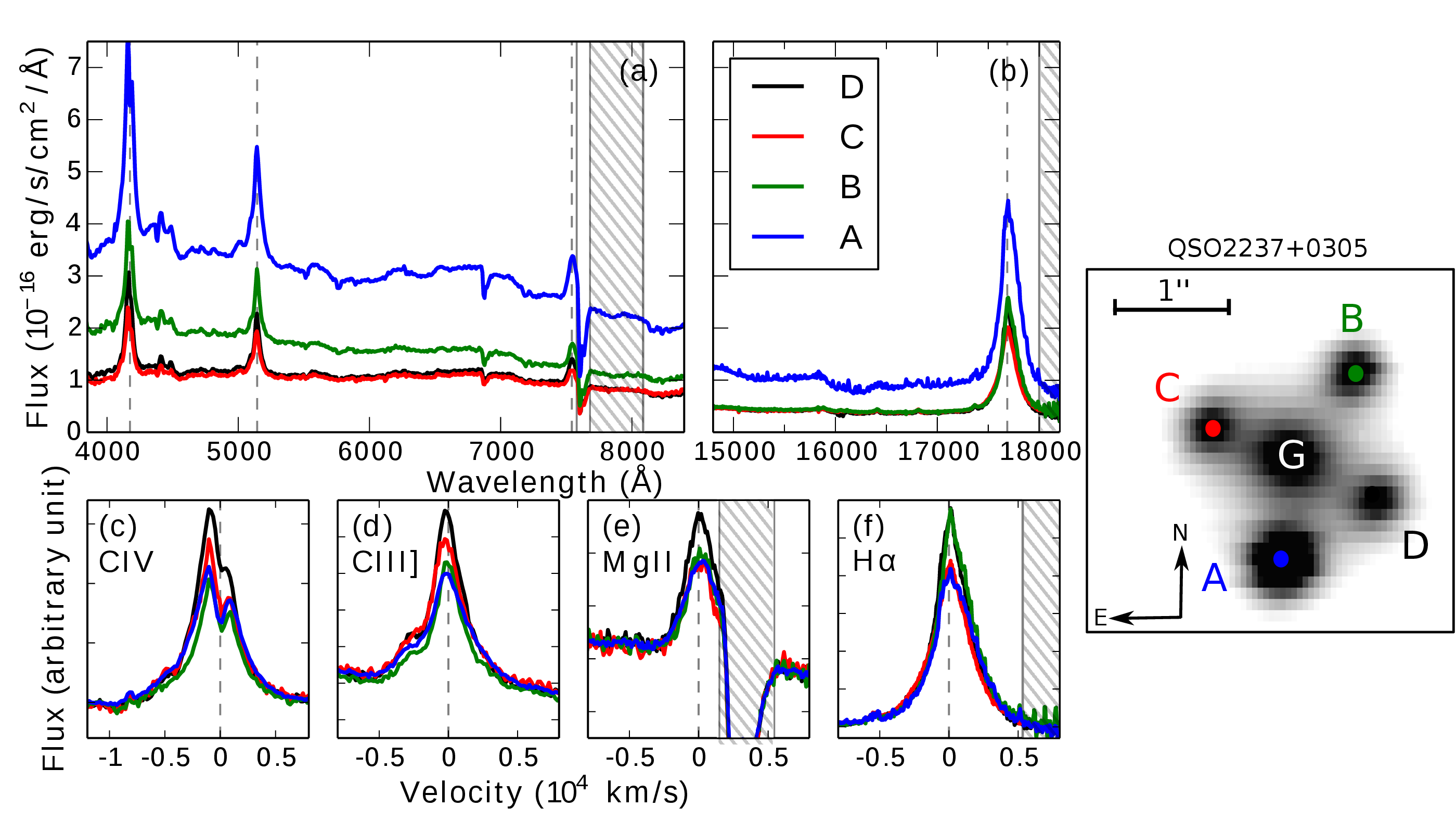}
\end{center}
\caption{(a) Visible spectra of the four images of QSO2237+0305 acquired on October 11 2005. (b) Mean near-infrared spectra obtained by combining all the valid spectra secured in October 2005. A typical SINFONI image of the QSO2237+0305 system, obtained by integrating the infrared flux over wavelength, is illustrated in the right panel. (c), (d), (e), and (f) Zoom on the CIV, CIII], MgII and H$\alpha$ line profiles, plotted vs the Doppler shift relative to the line laboratory wavelength redshifted to the quasar rest-frame. The spectra were rescaled so that their continua superimpose in the neighborhood of the lines. Vertical dotted lines indicate the positions of the lines in the quasar rest-frame. Hatched surfaces cover the wavelength ranges that suffer from important atmospheric absorption. The spectra are smoothed in wavelength using a 3-pixel wide median filter.}
\label{fig:show_spectra}
\end{figure*}

Fig.~\ref{fig:show_spectra} presents the spectra of the four lensed images of QSO2237+0305. Aside from the obvious brightness differences, the spectra of the four quasar images exhibit significant line profile differences, highlighted in the bottom panels of Fig.~\ref{fig:show_spectra} by rescaling the spectra so that their continua superimpose in the line's immediate surroundings. This emphasizes that (part of) the line emission does not behave like the underlying continuum emission. Since galactic dust extinction affects continuum and line emission similarly, and in the absence of significant time delays between the quasar lensed images, gravitational microlensing stands as the only phenomenon able to affect differently the continuum emission coming from the accretion disk and the broad lines emitted by the (more) extended BLR \citep{2008Yonehara}. Compared with the nearly flat continuum spectrum observed in both images C and D, images A and B display continua with large slopes, owing to chromatic microlensing of the accretion disk \citep{2008Eigenbroda}. 

Prominent microlensing events have been observed in the optical light curves of QSO2237+0305 images A, B and C, built from the OGLE long-term monitoring that extends over a 12-year period (\citealp{2000Wozniak}, \citealp{2006Udalski}), whereas no evidence of microlensing has been reported for image D (e.g., \citealp{2008Eigenbroda}, \citealp{2015Mediavilla}). The long-term (2.2 years) spectroscopic monitoring of QSO2237+0305 carried out by \citet{2008Eigenbroda} between October 2004 and December 2006 supports image D as the less affected by microlensing. We accordingly assume that image D is not affected by microlensing and use its spectrum as the reference quasar spectrum.

Fig.~\ref{fig:show_spectra} shows that the largest differential microlensing effect through the velocity structure of the lines takes place in image A. We therefore focus on the microlensing effect that affects that image. In the following subsections, we study the effect of microlensing on the line profiles using the macro-micro decomposition (MmD) method and the narrowband technique, which have the advantage of making no assumption on the line profiles. The multi-component decomposition (MCD), used in previous microlensing studies (\citealp{2008Eigenbroda}, \citealp{2011Sluse}), does not appear to be well suited because this method only involves symmetric Gaussian functions that cannot reproduce the asymmetric magnification of the Balmer line profile caused by microlensing (see Sect.~\ref{sec:MmD_Ha}).

\subsection{Macro-micro decomposition (MmD)}
\label{sec:MmD}

We use the MmD method (\citealp{2007Sluse}, \citealp{2010Hutsemekers}, \citealp{2012Sluse}) to reveal the part of the broad emission line that is microlensed as the underlying continuum. This decomposition method interprets the spectral differences caused by microlensing between two lensed images under the hypothesis that microlensing affects more strongly the continuum than the emission line. This assumption seems reasonable considering that microlensing is size-dependent and that the BLR is (much) larger than the accretion disk (e.g., \citealp{2007Kaspi}, \citealp{2011Sluse}).

MmD assumes that quasar image spectra, here $F_{\text{D}}$ and $F_{\text{A}}$,  can be expressed as linear combinations of a spectral component that is both macrolensed\footnote{We name ``macrolensing'' the magnification of the quasar image due to the whole lensing galaxy.} and microlensed, $F_{M\mu}$, and a spectral component that is only macrolensed, $F_M$. In image A, $F_M$ is therefore magnified by $M$, the A/D macro-amplification ratio, and $F_{M\mu}$ by $M \mu$, $\mu$ being the additional magnification caused by microlensing.
\begin{eqnarray}
\label{eq:fa}
F_\text{A} & = & M F_M + M \mu F_{M\mu} \\
\label{eq:fd}
F_\text{D} & = & F_M + F_{M\mu}  \hspace{3mm}.
\end{eqnarray}
By inverting Equations \eqref{eq:fa} and \eqref{eq:fd}, we can infer $F_M$ and $F_{M\mu}$ from the spectra of images A and D.

Following \citet{2011Sluse}, we fix the macro-amplification ratio to $M=1$, in agreement with the A/D$=1.00 \pm 0.10$ flux ratio measured in the mid-infrared \citep{2000Agol}, which is supposed to be unaffected by microlensing, and in agreement with macro-model expectations \citep{1998Schmidt}. The value of the microlensing factor, $\mu$, is determined independently in the continuum adjacent to each broad line, so that the $F_{M\mu}$ component includes the whole underlying continuum spectrum. 

We emphasize as a caveat that MmD, at a given Doppler shift, cannot disentangle the case where a part of the line flux is magnified like the continuum while the other part is not microlensed at all, from the case where the line flux is microlensed but less magnified than the continuum, i.e., $1<\mu^{line}<\mu^{continuum}$ (see appendix A of \citealp{2010Hutsemekers}). Nevertheless, when the line flux is entirely included in $F_{M\mu}$ (respectively in $F_M$) in some Doppler shift interval, it unambiguously indicates that this part of the line profile is magnified like the continuum (resp. not microlensed).



\begin{figure}
\centering
\includegraphics[width=0.5\textwidth]{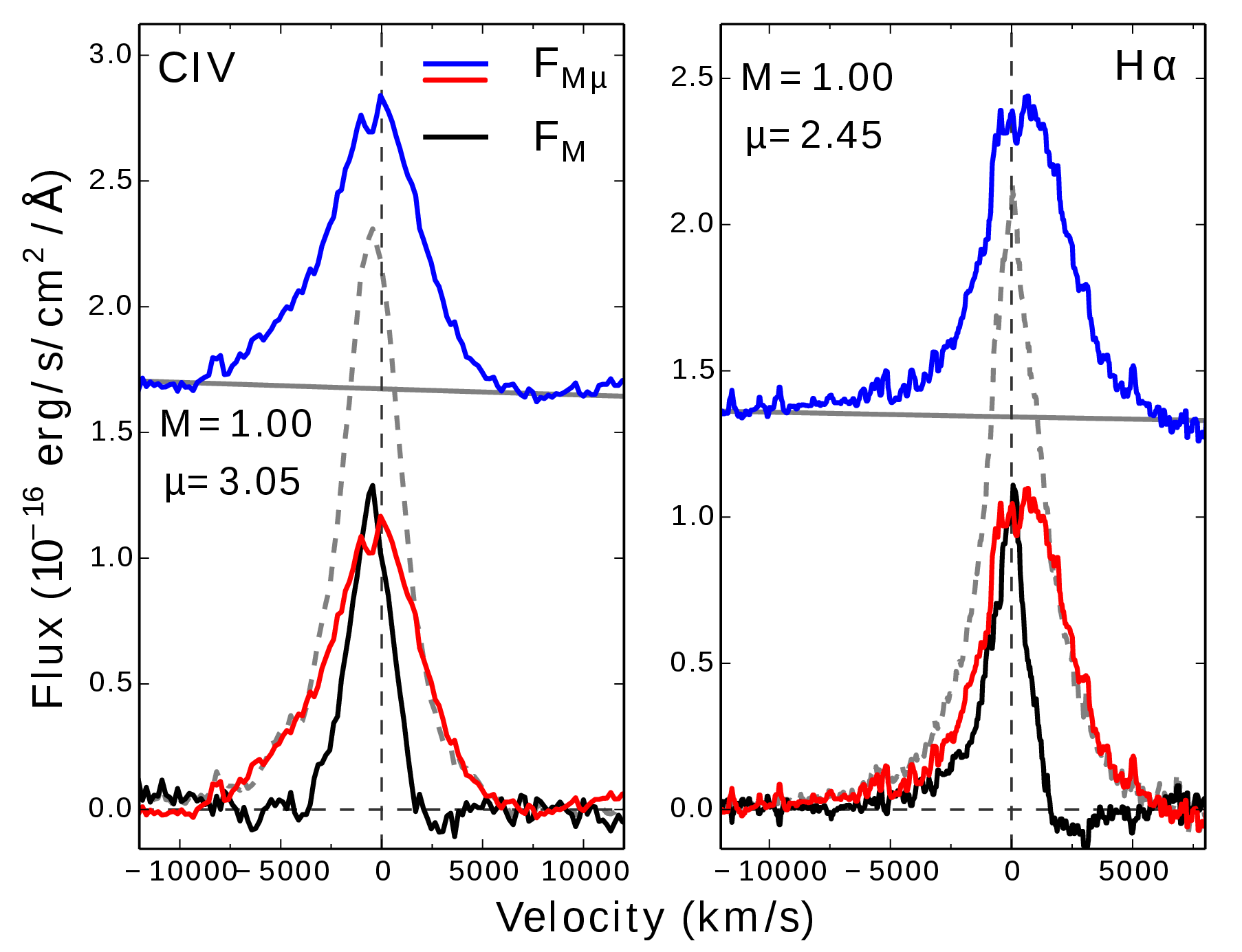}
\caption{The spectrum of quasar image A is decomposed into a microlensed component $F_{M\mu}$ (solid blue line) and a component $F_M$ (solid black line) which is supposed to be unaffected by microlensing. The microlensed part of the line profile, plotted with a solid red line, is obtained by subtracting the continuum from the microlensed component. Continuum level is indicated by a solid grey line. For clarity, the microlensed spectrum (blue line) is shifted up by $0.5 \hspace{1mm} 10^{-16}$ and $1.0 \hspace{1mm} 10^{-16}$ in the left and right panels respectively. CIV and H$\alpha$ line profiles observed in image D, which is supposed to be unaffected by microlensing, are plotted by a dashed gray line. The CIV line profile was corrected for the intervening absorption, clearly visible in Fig.~\ref{fig:show_spectra} (c), before applying MmD. The macro-amplification ratio, $M$, and the additional magnification caused by microlensing, $\mu$, determined for each line are indicated. Wavelengths are expressed in the quasar reference frame and converted into Doppler velocity shift, relative to the line laboratory wavelength.}
\label{fig:MmD_AD}
\end{figure}

The decompositions of the CIV and H$\alpha$ line profiles are illustrated in Fig.~\ref{fig:MmD_AD}. We discarded the CIII] and MgII line profiles from our analysis. The blue wing of CIII] is indeed blended with AlIII and SiIII] lines and MgII suffers from important atmospheric absorption, preventing any detailed analysis. The high-ionization CIV line and the low-ionization H$\alpha$ line show strikingly different MmD decompositions. This visible dissimilarity is asserted by the remarkably different shifts between the centroids of the $F_M$ and $F_{M\mu} $ components observed in the CIV and H$\alpha$ lines. The Doppler velocity shift of the line profile centroid, $v_{centroid}$ is computed as
\begin{equation}
v_{centroid} = \frac{\int F(v) \, v \, dv}{\int F(v) \, dv}
\label{eq:vcentroid}
\end{equation}
where $v$ is the Doppler velocity shift relative to the line laboratory wavelength and $F$ can be either $F_M$, $F_{M\mu}$ or $F_D$.

\subsubsection{The CIV high-ionization line}
\label{sec:MmD_CIV}

The CIV line profile is decomposed into a so-called narrow component seen in $F_M$, and a broad microlensed component included in $F_{M\mu}$, which is magnified by $3.05$ relatively to $F_M$ (left panel of Fig.~\ref{fig:MmD_AD}). This decomposition looks noticeably symmetric. The microlensed component contains a large part of the emission line, revealing that the large-amplitude microlensing effect which magnifies the UV continuum by $\mu^{continuum}=3.05$ also affects a large part of the high-ionization emission line.

$F_{M\mu}$ includes the whole line flux emitted at velocities $v<\unit{-4000}{\kilo\metre\per\second}$ and $v>\unit{2000}{\kilo\metre\per\second}$, which points out that both the blueshifted and redshifted high-velocity parts of the CIV line profile are magnified like the continuum. On the other hand, $F_M$ holds a large fraction of the line core. The core of the CIV line is definitely less microlensed than the underlying continuum, but we cannot distinguish whether microlensing highly magnifies a fraction of the emission line core or whether it magnifies the whole core emission less than the continuum. As noticed by \citet{2011Sluse}, the $F_M$ component has a FWHM that is larger than $\unit{1500}{\kilo\meter\second^{-1}}$ and, as a consequence, cannot be ascribed to the narrow emission line region.

A similar decomposition of the CIV line profile was obtained by \citet{2011Sluse} except that, in the present decomposition, we corrected CIV for the intervening absorption\footnote{The CIV absorption system has been identified as (at least) two clouds on the quasar line-of-sight, likely located in the quasar host galaxy (\citealp{1990Hintzen}, \citealp{1991YeeDeRobertis}). It is hence most probably not related to the quasar.} and subtracted blended iron emission prior to MmD. We used the same multi-component decomposition (MCD) as those authors to subtract an empirical iron pseudo-continuum template from the quasar image spectra. As in \citet{2011Sluse}, the detailed shape of the CIV line profile is reproduced by a sum of three Gaussian components: an absorption component, and a broad and a very broad emission component. We corrected the CIV line for absorption by adding the Gaussian absorption profile fitted with MCD. 

The centroid of the $F_{M\mu}$ component is blueshifted by $v_{centroid}^{CIV,F_{M\mu}} = \unit{-732 \pm 52}{\kilo\metre\per\second}$ and the centroid of the $F_M$ component by $v_{centroid}^{CIV,F_{M}} =\unit{-773 \pm 89}{\kilo\metre\per\second}$.\footnote{The centroids of the CIV broad emission line in the spectrum of image D, and of the microlensed and macrolensed parts of the CIV line profile, are computed using Eq.~\eqref{eq:vcentroid}, respectively with $F= F_D,$ $F_{M\mu}$ and $F_{M}$. $F_D$ and $F_{M\mu}$ are integrated over the $\unit{[-10000,10000]}{\kilo\metre\per\second}$ velocity range and $F_M$ is integrated over the $\unit{[-6000,6000]}{\kilo\metre\per\second}$ interval. When integrated over the same velocity range as $F_D$ and $F_{M\mu}$, $v_{centroid}^{CIV,F_{M}} =\unit{-908 \pm 173}{\kilo\metre\per\second}$}. The absence of significant relative shift between the $F_M$ and $F_{M\mu}$ components of the CIV line confirms the decomposition symmetry. Besides, the Doppler shifted centroids of the $F_M$ and $F_{M\mu}$ components are found to be compatible with the blueshift of the CIV broad emission line in the spectrum of image D, $v_{centroid}^{CIV,F_{M\mu}+F_M} =\unit{-798 \pm 50}{\kilo\metre\per\second}$. An apparent symmetric trend about the line center was also found by \citet{2011ODowd} in the B/A CIV line flux ratio, using spectroscopic observations of QSO2237+0305 acquired on June 27th, 2006, with the GMOS instrument on the Gemini-South telescope.

\subsubsection{The H$\alpha$ low-ionization line}
\label{sec:MmD_Ha}

As in the CIV line, a large part of the low-ionization emission line is magnified by a large-amplitude microlensing effect. However, contrary to the carbon line, the MmD decomposes the H$\alpha$ line into asymmetric components (right panel of Fig.~\ref{fig:MmD_AD}): while the red wing of the Balmer line profile is entirely included in $F_{M\mu}$ and is thus magnified like the continuum by $\mu^{line}=\mu^{continuum}=2.45$, the line core and the blue wing are partly contained in both $F_M$ and $F_{M\mu}$, which suggests a smaller amount of microlensing in that part of the line profile.

The centroid of the H$\alpha$ line in the spectrum of the image D is found to be compatible with a null Doppler shift, $v_{centroid}^{H\alpha,F_{M\mu}+F_M} =\unit{-141 \pm 136}{\kilo\metre\per\second}$, but we measure a significant centroid redshift of the $F_{M\mu}$ component of the H$\alpha$ line, $v_{centroid}^{H\alpha,F_{M\mu}} =\unit{126 \pm 29}{\kilo\metre\per\second}$, and an important blueshift of the centroid of the $F_{M}$ component, $v_{centroid}^{H\alpha,F_{M}} =\unit{-783 \pm 17}{\kilo\metre\per\second}$.\footnote{The centroid of the H$\alpha$ broad line profile and the centroid of its microlensed part are obtained by integrating $F_D$ and $F_{M\mu}$ over the $\unit{[-10000,8500]}{\kilo\metre\per\second}$ interval, using Eq.~\eqref{eq:vcentroid}. The centroid of the $F_M$ component is computed in the $\unit{[-8000,3000]}{\kilo\metre\per\second}$ velocity range.} The important relative shift between $F_M$ and $F_{M\mu}$ centroids confirms the asymmetric effect of microlensing in the H$\alpha$ low-ionization line.

\begin{table*}[t]
\caption{Narrowband A/D flux ratios computed in 7 velocity slices sampling the H$\alpha$ and CIV broad emission lines, along with 1$\sigma$-errors. CIV$^*$ column indicates flux ratios computed in slices defined with respect to the centroid of CIV macrolensed-only component, i.e., blueshifted by $\unit{-797}{\kilo\metre\per\second}$ compared with the CIV laboratory wavelength (sect.~\ref{sec:MmD_CIV}).}
\label{tab:narrowband_ADratios}
\centering
\begin{tabular}{ccccc}
\hline
\multicolumn{5}{c}{Narrowband A/D line flux ratios} \\
\hline
Band & Velocity Slice $(\unit{}{\kilo\meter\per\second})$ & H$\alpha$ & CIV & CIV$^*$ \\
\hline
B3 & $[-10000,-6000]$ &  $2.83 \pm 0.05$ & $4.43 \pm 0.38$ & $4.05 \pm 0.43$ \\
B2 & $[-6000,-4000]$ &  $2.23 \pm 0.02$ & $3.41 \pm 0.13$ & $3.85 \pm 0.21$ \\
B1 & $[-4000,-2000]$ &  $1.99 \pm 0.01$ & $2.77 \pm 0.05$ & $3.13 \pm 0.08$ \\
C & $[-1000,1000]$ &  $1.84 \pm 0.01$ & $2.15 \pm 0.02$ & $2.10 \pm 0.02$ \\
R1 & $[2000,4000]$ &  $2.74 \pm 0.01$ & $3.76 \pm 0.12$ & $3.27 \pm 0.07$ \\
R2 & $[4000,6000]$ &  $3.07 \pm 3.78$ & $4.23 \pm 0.39$ & $4.02 \pm 0.23$ \\
R3 & $[6000,8400]$ &  $1.98 \pm 3.07$ & $/$ & $/$ \\
\hline
\end{tabular}
\end{table*}

\subsection{Narrowband measurements}
\label{sec:NarrowBand}

To achieve a better understanding of the wing/core and red/blue microlensing effects affecting respectively the CIV and H$\alpha$ broad line profiles, the broad emission lines are sliced in velocity and the A/D line flux ratio is computed in each narrowband. This technique highlights differential microlensing throughout the line profile as variations of the line flux ratio.

We divide the broad line profiles into seven velocity slices. The line flux of quasar images A and D is integrated over each wavelength slice and its error is obtained by propagation. The line flux ratios computed at velocities larger than $\unit{6000}{\kilo\meter\per\second}$ are affected by very large errors because the line flux of image D reaches zero in that range. Velocity slices and A/D narrowband flux ratios computed for H$\alpha$ and CIV are listed in Table~\ref{tab:narrowband_ADratios}.

The A/D flux ratio estimated in the intermediate-velocity red wing (R1) of H$\alpha$ is about 25\% higher than in the intermediate-velocity  blue wing (B1, B2), in agreement with the larger magnification of the H$\alpha$ red wing unveiled by MmD (sect.~\ref{sec:MmD_Ha}). The flux ratio computed in the high-velocity part of the H$\alpha$ blue wing (B3) is however consistent with the red wing flux ratio, which suggests that highly blueshifted H$\alpha$ line emission is magnified like the red wing, and hence like the continuum. 

Consistent with the larger microlensing effect in the high-velocity part of the CIV wings that is emphasized by MmD (sect.~\ref{sec:MmD_CIV}), the A/D line flux ratios are found to increase significantly at high velocity in both the blue and red wings of the CIV line. Narrowband flux ratios reveal that the high-velocity part of CIV blue and red wings are twice as magnified by microlensing than the line core.

Yet, the flux ratio computed in the blueshifted part of CIV line profile (B1, B2) is $\sim 25 \%$ smaller than in the corresponding redshifted interval (R1, R2). This red/blue asymmetry found in CIV flux ratios could be due to the global blueshift of the line \citep{2011Sluse}. It indeed vanishes when Doppler shifted velocity bands are defined with respect to the centroid of the CIV line (CIV* in Table~\ref{tab:narrowband_ADratios}).

A minimum line flux ratio is measured in line core (C) for both the H$\alpha$ and CIV lines, which suggests the existence of a narrow unmicrolensed component, or at least that the microlensing effect is the weakest in the low-velocity part of both the high- and low-ionization broad lines.

\section{Discussion}
\label{sec:discussion}


\subsection{Constraints on the BLR structure}
\label{sec:disc-BLRconstrain} 

We found very dissimilar microlensing effects in the high-ionization carbon line and in the low-ionization Balmer line. The CIV broad emission line indeed appears affected by a wing/core microlensing effect while a red/blue effect is detected in the H$\alpha$ line. Given that high- and low-ionization regions are magnified by the same caustic structure, the different microlensing effects observed through the CIV and H$\alpha$ lines, especially at intermediate velocities, must arise from different geometry and/or kinematics of the high- and low-ionization regions.

\subsubsection{High-ionization region}
\label{sec:disc-highioConstrain}

The part of the caustic structure sampled by an emission region determines the magnification it experiences. For a given position on the magnification pattern, the resulting magnification is thus directly related to the source size and/or the distance between the source and the caustic(s). Significant magnification is expected for sources smaller than the Einstein radius of the system, i.e., about $50$\,lt-days \citep{1990Wambsganss}.

The part of the emission line which is strongly microlensed like the continuum spectrum must then arise from an emitting region significantly smaller than the Einstein radius. The size of the accretion disk of QSO2237+0305 has indeed been inferred to be smaller than 10\,lt-days (e.g., \citealp{2004Kochanek}, \citealp{2008Eigenbrodb}, \citealp{2010PoindexterKochanek}). This is also in agreement with the $20$\,lt-day size determined by \citet{2011Sluse} for the broadest Gaussian component of the CIV line, which includes the highly Doppler-shifted line flux. Hence, the high projected velocities of the high-ionization region velocity field are expected to concentrate in (a) compact region(s), which is/are likely to be co-spatial in projection with the continuum source, so that they sample similar region(s) of the caustic pattern.

On the other hand, all or part of the core of the CIV high-ionization line originates from a less magnified emission region. \citet{2011ODowd} interpreted the smaller magnification of the low-velocity part of the carbon line as low projected velocities that coincide with a low-magnification area of the caustic structure (Fig.~\ref{fig:mumap}). This can occur if the low-velocity part of the high-ionization region is sufficiently distant from the strongly-magnifying caustic(s) or sufficiently large to sample the low-magnification region(s) located farther away from the caustic(s). The former case implies that the low velocities are spatially separated from the high velocities and from the central accretion disk, in projection. 

In addition, the symmetric amplification of the redshifted and blueshifted parts of the CIV line profile requests that corresponding receding and approaching velocities sample similar regions of the caustic pattern, so that they undergo comparable magnification. This condition is trivially fulfilled when the receding and approaching parts of the velocity field are co-spatial in projection. Velocity fields with receding and approaching velocities spatially separated in projection are not completely ruled out but seem less likely given that caustic structures are highly asymmetric, so that fine-tuning would be necessary.

Beside the constraints set on the geometry and kinematics of the CIV emission region by microlensing, a valid model of the high-ionization region must generate significant blueshift of the line profile, such as the $\unit{-800}{\kilo\metre\per\second}$ centroid blueshift measured for the CIV line in the spectrum  of image D. 

\subsubsection{Low-ionization region}
\label{sec:disc-lowioConstrain}

MmD suggests that the redshifted wing of the H$\alpha$ line is microlensed like the underlying optical continuum, and thus likely comes from a compact source lying close to the accretion disk to sample neighboring regions of the caustic pattern. 
On the contrary, the weaker magnification measured in the blueshifted and low-velocity parts of the line profile indicates that these parts of the profile are emitted by regions spatially extended and/or separated from the continuum source. 

The red/blue dichotomy of the microlensing magnification measured through the H$\alpha$ line implies that the projected receding and approaching parts of the velocity field sample different areas of the caustic pattern. This can occur either when approaching velocities cover an extended region while receding velocities concentrate in a compact region in projection, or when projected receding and approaching parts of the velocity field are spatially separated from each other. The latter scenario is encountered in various BLR models, in particular in rotating velocity fields but also in radially expanding equatorial wind seen face-on.

Besides, the narrowband flux ratios suggest that the highly blueshifted part of the H$\alpha$ line undergoes a magnification comparable to the red wing (Sect.~\ref{sec:NarrowBand}). The high-velocity part of the H$\alpha$ blue wing could therefore come from a compact region, likely located close to the emission region of the highly redshifted part of the line profile and the continuum source seen in projection.

\subsection{Comparison with BLR models}
\label{sec:disc-BLRmodel}

The velocity and spatial structures of the BLR have been investigated for decades without reaching any consensus. Many models, involving radial outflow, inflow and/or gravitationally-dominated motions, with various geometries, are discussed in the literature. Those models generally reproduce a subset of the observed broad emission line properties. There is accumulating evidence suggesting that the BLR contains multiple components with different kinematics and/or geometrical properties, to reproduce the complex line profiles exhibited by the broad emission lines (e.g.,  \citealp{2004Popovic}, \citealp{2011Richards}, \citealp{2013GaskellGoosmann}, \citealp{2009Denney}) or to account for the different properties of high- and low-ionization lines (e.g.,  \citealp{2004BaskinLaor}, \citealp{2007Shang}).

In the following subsections, we confront the constraints on the CIV and H$\alpha$ emitting regions derived from microlensing (Sect.~\ref{sec:disc-BLRconstrain}) to various BLR models described in the literature. Consistent models are then considered in the context of a global picture of the BLR (Sect.~\ref{sec:blr-ionization-structure}).

\subsubsection{Keplerian disk}
\label{sec:disk-model}

\textbf{Observational evidence -} From the current, widely accepted, paradigm that the tremendous amounts of energy radiated by quasars are produced by a supermassive black hole fed by an accretion disk, follows the idea of a flat, gravitationally-dominated, BLR. Disk models are supported by the detection of double-peaked Balmer lines in a few active galactic nuclei (e.g., \citealp{1988Perez}, \citealp{1994Eracleous}). Yet, single-peaked broad emission line profiles are usually observed, which does not preclude the line from forming in a disk, provided that the emitting disk axis is seen at low inclination or that a part of the line core comes from a physically and kinematically different region (e.g., \citealp{2004Popovic}, \citealp{1997MurrayChiang}). The line core that partly arises from an extended emission region can especially account for the microlensing magnification dropping in the low-velocity part of both the CIV and H$\alpha$ lines.  

\begin{figure*}
\centering
\includegraphics[width=0.75\textwidth]{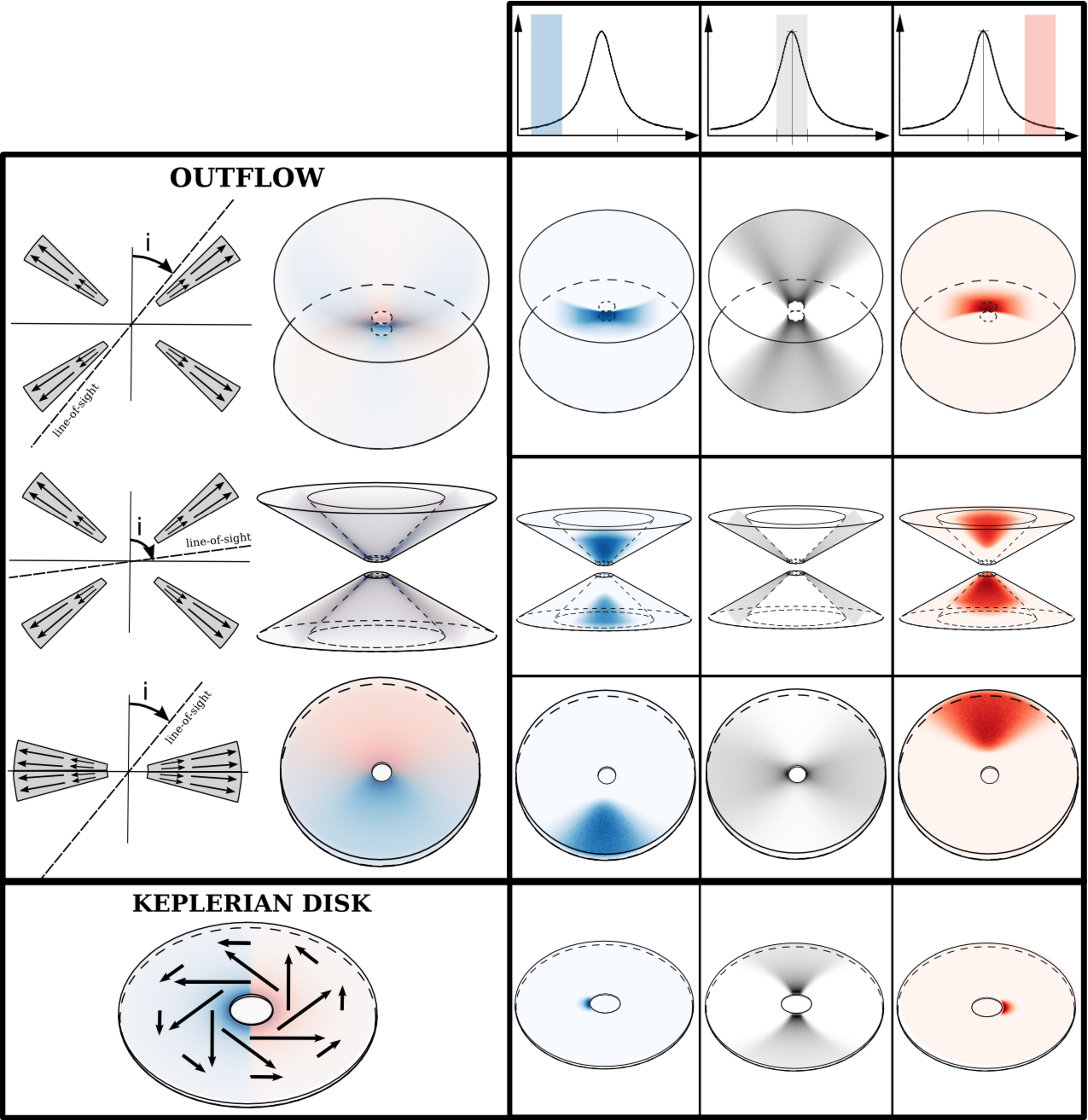}
\caption{Emission regions of the highly blueshifted, highly redshifted, and low-velocity line flux, illustrated for different BLR models: a polar wind with a conical shell geometry (rows 1 \& 2), an equatorial wind (row 3), and a Keplerian disk (row 4). The winds are radially accelerated. Their radial outflow is plotted in the left side of the outflow panel, in the plane of the observer line-of-sight for readability. The velocity fields are mapped with vectors, whose length is representative of the speed. Velocity slices are highlighted on the line profile drawn in the top of the figure, and the area of the BLR velocity field with matching projected velocities are plotted in the corresponding column. This figure is based on simulations of the emissivity of different BLR models, which will be the subject of a future paper.}
\label{fig:BLR_models}
\end{figure*}

\noindent
\textbf{Agreement with the wing/core effect in CIV?} In virial/Keplerian kinematics, the highest velocities are found at the innermost radii, close to the continuum source. The lower panels of Fig.~\ref{fig:BLR_models} indeed show that the highest-velocity parts of the line blue and red wings come from the most compact, innermost, regions, whereas the line core arises from a much larger region. This is consistent with the requirement that the core of the CIV line emerges from a more extended source than the high-velocity part of the wings (Sect.~\ref{sec:disc-highioConstrain}). However, in Keplerian disks, the approaching and receding parts of the velocity field are spatially separated, which cannot produce symmetric microlensing about the line center, except for specific caustic configurations. In addition, pure virial/Keplerian dynamics cannot produce a significant shift in the line profile, which argues against a virialized high-ionization region.



\noindent
\textbf{Agreement with red/blue effect in H\boldmath$\alpha$?} A Keplerian disk matches all the contraints on the low-ionization region derived from microlensing (Sect.~\ref{sec:disc-lowioConstrain}): the red/blue magnification dichotomy naturally arises from the spatial separation between negative and positive velocities and, since innermost radii rotate at highest velocity, regions emitting the most Doppler blueshifted and redshifted parts of the Balmer line are located close to each other and to the continuum source in projection. 

\subsubsection{Outflow}
\label{sec:outflows-model}

\textbf{Observational evidence -} An accelerating outflow is usually invoked to explain the existence of broad absorption line quasars. An outflow with an obscured receding part also constitutes a popular interpretation for high-ionization emission line shifts (e.g., \citealp{2002Richards}, \citealp{2004BaskinLaor}).\\
We investigate two outflow models: a polar outflow with a conical shell geometry and an equatorial outflow. QSO2237+0305 axis is supposed to be seen within $50\degree$ inclination \citep{2010PoindexterKochanek}. We assume that the line-of-sight does not cross the wind so that it is only seen in emission.

\textbf{A. Equatorial outflow}\\
As in the Keplerian disk, the approaching and receding parts of the radially expanding equatorial wind appear spatially separated in projection, when seen at intermediate inclination (Fig.~\ref{fig:BLR_models}, third row of panels). However, unlike the Keplerian disk, the extremely blueshifted and redshifted parts of the line come from regions that are located away from each other and from the continuum source, and are thus unlikely to be similarly magnified.

\noindent
\textbf{Agreement with the wing/core effect in CIV?} The comparable magnification of the negative and positive velocities, which are spatially separated in projection, along with the similar magnification of the continuum source and of the distant emission regions that produce the high-velocity part of the blue and red line wings, necessitate a properly oriented, symmetric, caustic pattern. Hence, an equatorial outflow could, but is unlikely to reproduce the wing/core effect in the CIV line.

\noindent
\textbf{Agreement with the red/blue effect in H\boldmath$\alpha$?} The spatial separation between the part of the equatorial outflow that approaches the observer and the part that recedes is consistent with the red/blue dichotomy of the magnification in the Balmer line. Yet, the large distance between the continuum source and the region from which the highly-redshifted part of the line profile originates, especially in an accelerating outflow, requires a fine-tuned caustic pattern to magnify these regions similarly, while less affecting the approaching part of the outflow. 

\textbf{B. Polar outflow}\\
In radial outflow, the highest approaching and receding velocities are located along the line of sight. Highly blueshifted and redshifted parts of the line wings then emerge from regions close to each other and to the continuum source in projection. As illustrated in the first row of panels of Fig.~\ref{fig:BLR_models}, the spatial separation between those regions emitting the high-velocity part of the line's blue and red wings is small compared with their extension, and they even appear partly co-spatial for line of sights that graze the wind conical shell. While high velocities concentrate in limited regions of the BLR, lower velocities cover larger regions in projection and are therefore less prone to microlensing magnification.
Although we address type 1 line of sights in this analysis, we note that corresponding positive and negative velocities appear to superimpose in projection when the polar wind is seen edge-on (see the second row of panels of Fig.~\ref{fig:BLR_models}), so that only symmetric magnification of the blue and red line wings can occur.\\
\noindent
\textbf{Agreement with the wing/core effect in CIV?} A polar wind seen at grazing incidence fulfills all the constraints on the geometry and kinematics of the CIV high-ionization region (Sect.~\ref{sec:disc-highioConstrain}). Symmetric magnification patterns are also exclusively observed when the polar wind is seen at sufficiently high inclination. Furthermore, partial obscuration of the receding part of the outflow by material in the equatorial plane provides a natural explanation for high-ionization line blueshifts.\\
\noindent
\textbf{Agreement with the red/blue effect in H\boldmath$\alpha$?} The polar outflow favors symmetric magnification of the line profile and thus cannot reproduce the microlensing signal observed in the H$\alpha$ broad line. 

\subsubsection{Inflow}
\label{sec:inflow-model}

\textbf{Observational evidence -} Velocity-resolved reverberation mapping measurements performed in a handful of AGNs have revealed that a variation of the red wing of high-ionization line profiles generally precedes a variation of the blue wing (e.g.,\citealp{1988Gaskell}, \citealp{1990CrenshawBlackwell}, \citealp{1996UlrichHorne}). \citet{2013GaskellGoosmann} interpret these results as the signature of inflowing material that scatters a fraction of the line emission, coming from an inner gravitationally-dominated region, into the observer's line-of-sight and towards bluer wavelengths. The addition of direct and scattered light results in a blueshifted line profile, skewed to the blue. The inner source of line emission must be significantly more compact than the surrounding scattering inflow in order to produce significant blueshift. The negligible blueshift of the low-ionization lines is then explained by their emission at larger radii from this inner source.

\noindent
\textbf{Agreement with the wing/core effect in CIV?} In this model, the most blueshifted line emission comes from an extended scattering region while the line red wing originates from the inner Keplerian disk, which is the source of direct line emission, close to the continuum source. Hence, the regions producing the most blueshifted and redshifted parts of the high-ionization line profile are located away from each other and likely sample distant areas of the caustic pattern, which disagrees with the constraints set by microlensing (Sect.~\ref{sec:disc-highioConstrain}). 

\noindent
\textbf{Agreement with the red/blue effect in H\boldmath$\alpha$?} Assuming that the inflowing material scatters a negligible part of the low-ionization emission line, the low-ionization region just behaves like a Keplerian disk and conclusions are identical to Sect.~\ref{sec:disk-model}.

\subsection{The global picture}
\label{sec:blr-ionization-structure}

Similarly to the results found for the lensed quasar HE0435-1223 \citep{2014Braibant}, a Keplerian disk nicely fits the constraints on the low-ionization region derived from the microlensing signal detected in the H$\alpha$ broad emission line (Sect.~\ref{sec:disk-model}). An equatorial wind cannot be completely ruled out, but involves fine-tuned caustic configurations to magnify the red wing of the H$\alpha$ line like the continuum without magnifying the blue wing.

On the other hand, the symmetric magnification of the CIV high-ionization line favors a polar wind (Sect.~\ref{sec:outflows-model}). A Keplerian disk or an equatorial wind can also match the microlensing effect detected in the carbon line, but they require a properly oriented symmetric caustic structure. Moreover, wind models advantageously provide a straightforward interpretation of the high-ionization line blueshifts. \citet{2011ODowd} interpreted the symmetric trend observed in the B/A magnification ratio through the CIV line as the signature of an outflow gravitationally-dominated at small radii. Disk-wind models or coexistence of gravitationally bounded and radiation driven CIV emission line regions have indeed been proposed by several authors (e.g., \citealp{1996ChiangMurray}, \citealp{2011Wang}, \citealp{2011Richards}). These combinations of outflow and virialized motions can reproduce both the line blueshift and the symmetric microlensing effect observed in the CIV line. Nevertheless, the polar outflow has the advantage of interpreting the wing/core microlensing pattern measured in CIV without need of a specific caustic pattern and conforms to the requirement that the geometry and/or kinematics of the high-ionization region differ from the low-ionization region, best modeled by a Keplerian disk (Sect.~\ref{sec:disc-BLRconstrain}).

Such a scenario has been proposed by \citet{2004BaskinLaor}, who explored the relation between the H$\beta$ and CIV lines. Those authors suggest that  a large relative accretion rate, $L/L_{Edd}$, could drive a lower density outflow that would contribute mostly to high-ionization lines while low-ionization lines might originate from denser lower-ionization gas lying deeper within the BLR structure, less affected by radiation pressure and mostly dominated by gravity.




\section{Conclusion}
\label{sec:conclusions}

We found that gravitational microlensing magnifies the blue and red wings of the H$\alpha$ low-ionization line differently, while it symmetrically magnifies the high-ionization CIV broad emission line. Since the high- and low-ionization regions are magnified by the same caustic structure, very different microlensing effects must be due to different kinematics and/or geometry of the high- and low-ionization gas. We argue that the asymmetric microlensing magnification observed through the Balmer line profile favors a Keplerian kinematics for the low-ionization region. In turn, the symmetric microlensing magnification in CIV favors a polar wind model for the high-ionization region. 

We emphasize that a similar red/blue microlensing magnification has been detected in the Balmer broad emission line of the image D of the HE0435-1223 gravitationally lensed quasar \citep{2014Braibant}, which makes the interpretations involving caustic fine-tuning less likely and further supports the Keplerian disk model for the low-ionization region. By contrast, the polar wind model is the only one that leads to comparable magnification of the very blueshifted and redshifted parts of the line profile when seen at intermediate inclination, whatever the magnifying caustic structure. Among the 18 quasars for which microlensing of the high-ionization lines is observed (\citealp{2004Richards}, \citealp{2012Sluse}, \citealp{2013Guerras}), only SDSS J1004+4112 shows a significant differential magnification of the blue and red wings in the CIV line profile ; if indeed caused by microlensing, this could constitute an important blow to high-ionized gas flowing as a wind. Further study of this object is badly needed. Still, most systems show no clear differential microlensing and/or a small amount of microlensing in their high-ionization line profiles and thus set no constraint on the high-ionization region geometry and kinematics.

In the previous paragraph, we implicitly assumed that all quasars, including the Einstein cross, are similarly constituted.  If so, BLR models can be statistically tested by studying large amplitude microlensing effects on the high- and low-ionization lines in other lensed systems. Moreover, better insight into the BLR structure can be achieved by observing the temporal evolution of the line profile distortions caused by microlensing, in particular because this enables us to rule out fine-tuned caustic configurations.

\begin{acknowledgements}
L. Braibant and D. Hutsem\'ekers acknowledge support of Belgian F.R.S.-FNRS. D. Sluse acknowledges support from a ``Back to Belgium'' grant from the Belgian Federal Science Policy (BELSPO). Support for T. Anguita is provided by proyecto FONDECYT 11130630 and the Ministry of Economy, Development, and Tourism’s Millennium Science Initiative through grant IC120009, awarded to The Millennium Institute of Astrophysics, MAS.
\end{acknowledgements}

\bibliographystyle{aa}
\bibliography{references}

\onecolumn

\begin{appendix}

\section{Observing conditions during near-infrared spectrum acquisition}
\begin{table*}[h!]
\begin{center}
\caption{Near-infrared observations of the lensed quasar, QSO2237+0305, with the SINFONI Integral Field spectrograph at ESO-VLT (program ID 076.B-0607, principal investigator: Courbin).}
\label{tab:sinfo_obs_conditions}
\begin{tabular}{|lcccc|}
\hline
Civil date & Time [UT] & Seeing[''] & Airmass & Sky \\
\hline
01-10-2005 & 1h51 & 0.64 & 1.162 & clear \\
01-10-2005 & 2h23 & 0.64 & 1.136 & clear \\
02-10-2005 & 0h43 & 0.84 & 1.299 & clear \\
02-10-2005 & 1h14 & 0.98 & 1.216 & clear \\
02-10-2005 & 1h54 & 0.64 & 1.155 & clear \\
14-10-2005 & 0h02 & 0.60 & 1.279 & clear \\
14-10-2005 & 0h34 & 0.61 & 1.203 & clear \\
14-10-2005 & 0h44 & 0.61 & 1.184 & clear \\
14-10-2005 & 1h28 & 0.59 & 1.138 & clear \\
15-10-2005 & 0h13 & 0.88 & 1.239 & clear \\
15-10-2005 & 0h45 & 0.90 & 1.177 & clear \\
15-10-2005 & 23h51 & 0.60 & 1.290 & clear \\
16-10-2005 & 0h22 & 0.54 & 1.210 & clear \\
17-10-2005 & 0h02 & 1.14 & 1.247 & not available \\
17-10-2005 & 0h33 & 0.94 & 1.183 & not available \\
18-10-2005 & 0h17 & 0.47 & 1.204 & clear \\
18-10-2005 & 0h49 & 0.47 & 1.157 & clear \\
19-10-2005 & 0h13 & 0.66 & 1.204 & clear \\
19-10-2005 & 0h45 & 0.57 & 1.157 & clear \\
23-10-2005 & 0h55 & 0.45 & 1.137 & clear \\
23-10-2005 & 1h26 & 0.51 & 1.134 & clear \\
23-10-2005 & 1h44 & 0.45 & 1.141 & clear \\
23-10-2005 & 2h16 & 0.43 & 1.174 & clear \\
24-10-2005 & 23h59 & 0.50 & 1.187 & almost clear \\
25-10-2005 & 0h30 & 0.41 & 1.148 & almost clear \\
25-10-2005 & 0h51 & 0.33 & 1.136 & almost clear \\
25-10-2005 & 1h23 & 0.33 & 1.135 & almost clear \\
26-10-2005 & 0h19 & 0.57 & 1.156 & clouds \\
26-10-2005 & 0h50 & 0.70 & 1.135 & clouds \\
28-10-2005 & 23h54 & 0.41 & 1.171 & clear \\
29-10-2005 & 0h26 & 0.57 & 1.140 & clear \\
29-10-2005 & 0h36 & 0.42 & 1.135 & clear \\
29-10-2005 & 1h08 & 0.41 & 1.135 & clear \\
29-10-2005 & 23h55 & 0.49 & 1.165 & clear \\
30-10-2005 & 0h27 & 0.41 & 1.138 & clear \\
30-10-2005 & 0h37 & 0.39 & 1.134 & clear \\
30-10-2005 & 1h09 & 0.45 & 1.137 & clear \\
30-10-2005 & 1h51 & 0.49 & 1.162 & clear \\
30-10-2005 & 2h22 & 0.49 & 1.239 & clear \\
30-10-2005 & 2h33 & 0.47 & 1.267 & clear \\
30-10-2005 & 3h05 & 0.55 & 1.377 & clear \\
\hline
\end{tabular}
\end{center}
\end{table*}

\end{appendix}

\end{document}